\documentclass[aps,prl,showpacs,twocolumn,superscriptaddress,showpacs,groupedaddress,nofootinbib]{revtex4-1}  

\usepackage{color}
\usepackage{graphicx}  
\usepackage{dcolumn}   
\usepackage{bm}        
\usepackage{amssymb}   
\usepackage{amsmath,amssymb,amsthm,mathrsfs,amsfonts,dsfont} 
\usepackage[colorlinks=true,citecolor=magenta, linkcolor=darkblue]{hyperref}
\usepackage[dvipsnames]{xcolor}
\definecolor{darkblue}{rgb}{0.0, 0.0, 0.55}

\pdfoutput=1

\newcommand{\myhref}[1]{\href{https://arxiv.org/abs/#1}{{\color{darkblue} #1}}}
\def\be{\begin{eqnarray}}
\def\ee{\end{eqnarray}}

\def\Vol{\operatorname{Vol}}

\def\d{{\rm d}}

\def\cO{\mathcal{O}}
\def\b{\bar}

\begin{document}


\title{
How $a$-type anomalies can depend on marginal couplings
}

\author{Christopher P.\ Herzog$^{1}$ and Itamar Shamir$^{2}$ \vspace{0.1cm}}
\affiliation{
$^{1}$
Mathematics Department, King's College London, \\
The Strand, London,  WC2R 2LS, UK
}
\affiliation{
$^{2}$
SISSA and INFN, Via Bonomea 265, 34136, Trieste, Italy
}


\fontsize{10pt}{11.7pt}\selectfont

\begin{abstract}
Even dimensional defects and boundaries in conformal field theory support type $a$ anomalies on their world-volume. We show that the one-point functions of marginal operators, in the presence of defects and boundaries, are anomalous, and that the Wess-Zumino consistency condition 
relates them to the derivative of the $a$-anomaly with respect to the marginal coupling. 
We also argue that the constant term $F$ for odd dimensional surfaces can depend on marginal parameters.

\end{abstract}
\maketitle

 Boundaries and defects in quantum field theory (QFT) play an important role in many recent developments in theoretical physics: D-branes, AdS/CFT, quantum entanglement in many-body systems, and topological insulators to name a few.  These examples suggest boundary and defect QFT is worth studying in its own right, admittedly a vast subject.  One target of opportunity is conformal field theory (CFT) which provides landmarks in the space of QFT more generally -- points where the beta functions cease to run and the theories become scale invariant. Additionally, CFTs are useful experimentally in describing certain condensed matter systems at their phase transitions. 

Some CFTs are of special interest because they belong to larger families, parametrized by a 
set of coupling constants for exactly marginal operators.  
The associated continuous space, with singularities, is called the conformal manifold, and admits a natural metric, induced from the two point function of these operators \cite{Zamolodchikov:1986gt}.  More refined structures than a metric can emerge as well, e.g.\ when the theory has supersymmetry.

It is natural to ask how various observables of CFT depend on marginal parameters. Some of the most interesting observables come from the sphere partition function. In even dimensions, the only scheme independent contribution to the sphere partition function, so called $a$ (or $c$ in 2d), is associated with the conformal anomaly,\footnote{%
It is traditional to include an extra sign $(-1)^{k+1}$ in the definition of $a$ (and presumably $F$), 
to guarantee positivity, which we here suppress for notational brevity.
} 
\begin{align} \label{Za}
\log Z 
=
\begin{cases}
2 a \log \Lambda \, , & d=2k \, ,
\\
-F \, , & d=2k+1 \, ,
\end{cases}
\end{align}
while in odd dimensions a constant term in the partition function called $F$ can be unambiguously defined. Both $a$ and $F$ are fundamental to our understanding of renormalization group flows. In two, three, and four dimensions (without boundary), for instance, it is known that these $a$ and $F$ quantities must decrease under renormalization group (RG) flow \cite{Zamolodchikov:1986gt,Komargodski:2011vj,Casini:2012ei}. The $a$-type surface anomaly that is the focus of this letter must decrease under boundary RG flow \cite{Jensen:2015swa}.  There are arguments that $a$ and $F$ quantities associated with hypersurfaces should also have a monotonicity property \cite{Kobayashi:2018lil,Casini:2018nym,Gaiotto:2014gha}.

Another way of characterizing the $a$-anomaly is through the trace of the stress tensor.
While scale invariance implies classically a vanishing trace $T^\mu_{\; \mu} = 0$, quantum effects on a curved manifold mean that $T^\mu_{\; \mu}$ can be proportional to a set of curvature invariants with special properties. 
We focus here on the Euler density $E_d$ -- the curvature invariant which integrates to the Euler characteristic on a compact manifold.  The $a$-anomaly is traditionally identified as the coefficient of the Euler density 
$T^\mu_{\; \mu} = a E_d  +\ldots$ where the ellipses denote other possible anomalous contributions to the trace \cite{Deser:1993yx}.  

It is known from general principles that both kinds of observables, $a$ in even dimensions or $F$ in odd, are constrained to be constant on the conformal manifold. Indeed, it follows from the Wess-Zumino consistency condition \cite{Wess:1971yu,Osborn:1991gm} that the $a$-anomaly must be independent of marginal parameters. In odd dimensions, $F$ must likewise be independent since the one-point function must vanish in the absence of a conformal anomaly 
\cite{Gerchkovitz:2014gta}.\footnote{%
With enough supersymmetry the constant term in even dimensions becomes well defined, in which case it depends non-trivially on marginal directions \cite{Gerchkovitz:2014gta,Gomis:2015yaa}. } 

The goal of this letter is to explore how this picture of $a$-anomalies changes for CFTs with boundaries or defects. 
For a $p$ dimensional boundary or defect, where $p$ is even, quite generally we can continue to identify an $a$-type anomaly associated with the defect, 
\begin{align} \label{Tmumu}
T^\mu_{\; \mu} = a  E_p \delta^{(q)}(z)+\ldots
\end{align}
where $z$ are coordinates specifying the $q = d-p$ directions transverse to the defect.  Equivalently, we can isolate the $a$-anomaly 
from a partition function. A $p$ dimensional planar defect in flat space can be mapped by a conformal transformation to 
an equatorial $S^p$ inside $S^d$
where $S^p$ is a $p$ dimensional sphere.  
In the special case $q=1$, we can treat the $S^{d-1}$ either as a defect inside $S^d$ or the boundary
of a hemisphere $HS^d$.

In what follows, we begin 
by discussing a scale anomaly in the one-point function  for the marginal operators ${\mathcal O}_I$, 
and in particular how this anomaly does not satisfy WZ consistency on its own.
We show how WZ consistency can be restored by letting the $a$-anomaly depend on the corresponding marginal couplings $\lambda^I$.  
 We then discuss the partition function on $S^d$.  In particular, we show how a scheme independent contribution to the partition function, which corresponds to the $a$-anomaly, depends on the marginal parameters $\lambda^I$.  As a check, we see that the dependence of the $S^d$ partition function on $\lambda^I$ matches the dependence that WZ consistency imposes on the $a$-anomaly.

\vskip 0.1in
\noindent
{\it One-Point Function Anomaly}

Consider a  defect CFT with a set of exactly marginal (bulk) operators $O_I$ and coupling constants $\lambda^I$. We write the partition function schematically as a Euclidean path integral
\be
Z = \int [\d \phi] \exp \left(-S[\phi] - \sum_I  \int_M \d^d x \sqrt g \, \lambda^I \cO_I \right) \ ,
\ee
and define the effective action as $W \equiv - \log Z$.
To investigate the dependence on $\lambda$ we consider the derivative of the effective action, which gives the integrated one-point function
\begin{align} \label{der_Z_I}
\partial_I W = \int_M \d^d x \sqrt g \,  \langle \mathcal{O}_I (x) \rangle.
\end{align}
In flat space, the one-point function of these marginal operators is fixed by conformal symmetry to the form \cite{McAvity:1993ue,McAvity:1995zd,Billo:2016cpy}
\be
\langle {\mathcal O}_I \rangle = \frac{f_I(\lambda)}{|z|^d} \ ,
\ee
where $|z|$ is the distance from the defect (note that $z$ is $q$ dimensional). 
The defect allows for a nonzero $f_I(\lambda)$ which in turn means that the partition function $W$
will in general depend on $\lambda$ as well.

We claim that there is a scale anomaly associated with the divergence in $\langle {\mathcal O}_I \rangle$ as $
z \to 0$.  
To understand where this anomaly is coming from, we ask whether the one-point function specifies a well-defined distribution. While this question may seem formal, it is in fact quite natural from the following point of view. The basic object of interest is the effective action $W$, viewed as a functional of the background coupling $\lambda(x)$ and the metric $g_{\mu\nu}(x)$ both promoted to background fields. For the effective function to be well-behaved for sufficiently nice background fields, the correlation functions of $\mathcal{O}$ as well as the stress-energy tensor must be well-defined distributions.

To approach this question consider an operator $\mathcal{O}_\Delta$ of arbitrary dimension $\Delta$, such that its one-point function is $\langle \mathcal{O}_\Delta \rangle \sim |z|^{-\Delta}$. Such distributions are known as homogeneous in the mathematical literature. From our point of view the homogeneity is of course a manifestation of conformal symmetry. It is a well known result (see for instance \cite{distr}) that $|z|^{-\Delta}$ can be analytically continued in $\Delta$ leading to a homogeneous distribution, except for poles at special values of $\Delta$. For these $\Delta$'s the distribution is made well-defined only at the cost of spoiling the homogeneity, i.e.\ introducing an anomaly. To find which values of $\Delta$ lead to an anomaly we can argue as follows. On dimensional grounds the anomaly takes the form 
\begin{align} \label{}
\Lambda \partial_\Lambda |z|^{-\Delta} \sim \left( \Box_q \right) ^{\frac{\Delta-q}{2}} \delta^{(q)}(z) \, .
\end{align}
Since the anomaly must be local $\Delta-q=2k$ where $k=0,1,2,\ldots$ except for $q=1$ where $\Delta-1=k$.%
\footnote{As a trivial application we can think of a point like defect such that $q=d$ and $|z|^{-\Delta}$ is now viewed as a two-point function of an operator of dimension $\delta = \Delta/2$. Such two-point functions are anomalous when $2\delta = d+2k$. 
For a marginal operator with $\delta=d$ the anomaly only exists in even dimensions. 
}

We now focus on the case $\Delta=d$. 
To derive the anomaly we follow the real space renormalization arguments of \cite{Freedman:1991tk}. 
Unfortunately, we need to treat the various codimension $q$ cases separately.  In the case $q=1$, we can use
the operator $\partial_z$ in the regularization procedure, while in higher codimension, it is more natural to use  $\Box_q$.
The result for $q>2$ can be written in generality, but the limit $q\to 2$ requires some additional care.

Let us start with the simplest case, $q=1$.  We can partially regularize the divergence in the one-point function
by introducing a scale $\Lambda$,
\begin{align} \label{}
z^{-d} = \begin{cases}\frac{(-1)^{d-1}}{(d-1)!}\partial_z^{d} \log (z\Lambda), & z>0 \\ 0, & z<0 \end{cases} \ .
\end{align}
 Performing a scale transformation $\Lambda \partial_\Lambda$ 
leads to the anomalous term 
\begin{align} \label{one_pt_anomaly}
\Lambda \partial_\Lambda z^{-d} = \frac{(-1)^{d-1}}{(d-1)!} \partial_n^{d-1} \delta(z) \, .
\end{align}
Here we use $\partial_n = n^\mu \partial_\mu$ for normal derivatives on the boundary.

For $q>1$, in contrast, the anomaly takes the form 
\begin{align} \label{anomalyq}
\Lambda \partial_\Lambda |z|^{-d} = c_{p,q} (\Box_q)^{\frac{p}{2}} \delta^{(q)}(z) \, ,
\end{align}
where now $p$ is restricted to be even. The constant $c_{p,q}$ is straightforward to work out in general, but our interest in what follows is the special case of surface defects for which $p=2$, and for which we find
\begin{align} \label{}
|z|^{-d} 
= 
\begin{cases} 
\frac{1}{8} \Box_2^2 \log^2 (|z| \Lambda)\, , & q=2 \, ,
\\
\frac{1}{2q(2-q)} \Box_q^2 \left( |z|^{2-q} \log (|z| \Lambda)\right)\, , & q>2 \, . 
\end{cases}
\end{align}
Thus
\be
\label{c2q}
c_{2,q} = 
\frac{\Vol(S^{q-1})}{2q}  \ ,
\ee
where $\Vol(S^{q-1}) = 2 \pi^{q/2} / \Gamma(\frac{q}{2})$ is the volume of a sphere of unit radius. 
Note there is a factor of two discrepancy between the $q \to 1$ limit of (\ref{c2q}) and the $d=3$ case of 
(\ref{one_pt_anomaly}), which stems from the fact that we treat the $q=1$ case as a boundary. 
In contrast, \eqref{c2q} contains the volume of a zero dimensional sphere with two points, reflecting a two sided defect rather than a boundary.

The existence of the anomalies (\ref{one_pt_anomaly}) and (\ref{anomalyq})
 means that the Weyl transformation of the
 effective action contains the boundary contribution
 \begin{align} \label{one_point_eff_action}
\Lambda \partial_\Lambda W = c_{p,q} \int \d^{p} x  \, f_I \, (\Box_{q})^{\frac{p}{2}} \lambda^I \, ,
\end{align}
written here in flat space. While in general the anomaly exists for even $p$,  in the special case $q=1$ the formula extends to odd $p$ as well.\footnote{In principle, some (but not all) of the derivatives can act on $f_I$ as well, and still reproduce the one-point anomaly. The difference will not matter for the case we study in this letter. (We discuss the uniqueness of the solution below.)}

To study the implications of this anomaly, we first need to understand the curved space generalization of (\ref{one_point_eff_action}).  
As usual, the details significantly depend on the dimension. We will focus here on surface defects, $p=2$.
Trading
$\Lambda \partial_\Lambda$ for a Weyl transformation $g_{\mu\nu} \to e^{2 \sigma} g_{\mu\nu}$, we can write \eqref{one_point_eff_action} more generally as
 \begin{align} \label{var_op_anomaly}
\delta W 
= 
c_{2,q} \int \d^2 x \sqrt \gamma \, \delta \sigma \, f_I(\lambda) \delta^{ij} n_i^\mu n_j^\nu \b D_\mu \b D_\nu \lambda^I \,, 
\end{align}
where $\gamma$ is the induced metric on the boundary and the vectors $n_i$ are unit normalized and normal to the surface. We use $\b D$ for the bulk connection and $D$ for the induced connection.\footnote{One can easily include a connection $\Gamma_{IJ}^K$ in the space of couplings. The difference is Weyl invariant by itself and hence doesn't affect our result. We thank A.~Schwimmer for discussion on this point.} 

This term (\ref{var_op_anomaly}) cannot stand by itself since it does not satisfy the WZ consistency condition \cite{Wess:1971yu}. This condition follows from the Abelian nature of the Weyl symmetry and requires that two Weyl transformations commute, namely $[\delta, \delta'] W=0$. Note that under a Weyl rescaling, the normal changes by $\delta n^\mu = - \delta \sigma \, n^\mu$ and the (bulk) connection by
\begin{align} \label{}
&n^\mu \cdot n^{\nu} \delta \b \Gamma_{\mu\nu}^{\alpha} = (2-q) n^\alpha \cdot \partial_{n} \delta \sigma  - q \gamma^{\alpha\beta} \partial_\beta \delta \sigma   \, .
\end{align}
The only non-trivial contribution to the action of $\delta'$ on \eqref{var_op_anomaly} comes from the connection
\begin{align} \label{}
\delta' \delta W  = 
c_{2,q} \int \d^2 x & \sqrt \gamma \, \delta \sigma \, f_I \left(q \partial_\alpha \delta \sigma' \gamma^{\alpha\beta} \partial_\beta \lambda^I  \right. \nonumber \\
& \left. + (q-2) \partial_n \delta \sigma'  \cdot \partial_n \lambda^I \right) \, , 
\end{align}
which is not symmetric exchanging $\delta \sigma$ for $\delta \sigma'$.

Given that $\delta W$ must be a local functional of the background metric and couplings, we can consider adding the following two terms
\begin{align} \label{delta_sigmaW}
\int \d^2 x \sqrt \gamma \, \delta \sigma \, \left( 
-\frac{a(\lambda)}{4 \pi} 
R + b_I(\lambda) K \cdot \partial_n \lambda^I \right) \ .
\end{align}
Here $R$ is the 2d Ricci scalar and $K^i$ is the trace of one of the extrinsic curvatures, $i = 1, \ldots, q$. 
The quantity $a(\lambda)$ is the $a$-anomaly of (\ref{Tmumu}).
Under a Weyl rescaling  
\begin{align} \label{}
\delta R &= -2 \delta \sigma R - 2 D^2 \delta \sigma \, , \nonumber \\
\delta K^i &= -\delta \sigma K^i + 2\partial_{n_i} \delta \sigma \, . 
\end{align}
Applying the consistency condition we get the relations
\begin{align} \label{farel}
\partial_I a =  2 \pi  q c_{2,q}f_I  \, , 
\qquad 
 2b_I =(2-q) c_{2,q}  f_I \, .
\end{align}
We see that the coefficient of the boundary $a$-anomaly is determined by the one-point function anomaly up to an integration constant.
Note that there is no way to cancel the variation of $R$ (the $a$-anomaly) intrinsically in 2d, unless $a$ is constant.
This corresponds to the standard result that $a$ cannot depend on marginal parameters. Indeed, it is still true that $a$ cannot depend on defect marginal parameter associated with operators localized on the defect. Here, the extrinsic contribution coming from the one-point anomaly is crucial for avoiding this fate. 

The solution of the WZ conditions we found above is unique only up to terms which satisfy the WZ condition independently, which means that their coefficient cannot be fixed by such considerations. A different solution is given by considering the bulk Laplacian $\b D^2$. On the boundary it can be written 
\begin{align} \label{}
\b D^2 = D^2 + K\cdot \partial_n + \delta^{ij} n_i^\mu n_j^\mu \b D_\mu \b D_\nu \, .
\end{align}
In two dimensions, the (induced) Laplacian $D^2$ transforms homogeneously under a Weyl transformation and thus we can substitute \eqref{var_op_anomaly} for $\frac{1}{2} \int \d^2 x \sqrt{\gamma}\, \delta \sigma f_I \b D^2 \lambda^I$, provided we also shift $b_I \to b_I - \frac{1}{2} f_I$.

\vskip 0.1in
\noindent
{\it Partition Function}

Let us now consider the $S^d$ partition function.\footnote{%
Another nice geometry conformal to flat space is a product of hyperbolic space and a sphere, $H_{p+1} \times S^{q-1}$
\cite{Kapustin:2005py,Rodriguez-Gomez:2017kxf}, where the defect now lives on the boundary of $H_{p+1}$.  
As it is more difficult to deal with counter-terms in $H_{p+1} \times S^{q-1}$, we focus on $S^d$ instead. See \cite{Rodriguez-Gomez:2017aca} for a related geometry with a ball instead of hyperbolic space. 
} 
We write the metric in a particular way
\be
\d s^2 = \d \theta^2 + \sin^2 \theta \, \d \Omega_p^2 + \cos^2 \theta\,  \d \Omega_{q-1}^2 \ ,
\ee
where $\d \Omega_p^2$ is the line element on a sphere of unit radius. 
(For convenience, we set the radii of curvature to one.)  The defect is chosen to lie on the $S^p$ defined by
$\theta = \frac{\pi}{2}$. This configuration is conformally related to a planar defect in flat space. 
Through the Weyl transformation, we find in this coordinate system that 
\be
\langle {\mathcal O}_I \rangle_{S^d} = \frac{f_I(\lambda)}{ \cos^d \theta} \ .
\ee
The derivative of the partition function takes a divergent form
\begin{align} \label{partialWinit}
\partial_I W_{p,q} = f_I \, \Vol(S^{p}) \Vol(S^{q-1}) \int_0^\frac{\pi}{2} \frac{\sin^p \theta}{\cos^{p+1} \theta} \d \theta .
\end{align}

We can regularize the integral (\ref{partialW}) in various ways.  Perhaps the simplest is dimensional regularization
which yields the result
\be
\label{partialW}
\partial_I W_{p,q} = -  f_I \,  \frac{\Vol(S^{p+1}) \Vol(S^{q-1})}{2 \sin \left(\frac{\pi p}{2} \right)} \ .
\ee
There is a divergence for even values of $p$.  
This log divergence is symptomatic of the fact that the defect is even 
dimensional and may support a scale anomaly.

To see more clearly how the result (\ref{partialW}) may or may not depend on the regularization scheme, it is useful to evaluate the volume integral using a different scheme, a cut-off prescription where $0 < \theta < \frac{\pi}{2} - \delta$.  In this alternate regularization scheme, we find
in dimensions $p=1$, 2 and 3 that 
\be
\frac{\partial_I W_{1,q}}{f_I \Vol(S^{q-1})} &=& \frac{2\pi}{\delta} - 2 \pi + O(\delta) \ , \\
\frac{\partial_I W_{2,q}}{f_I \Vol(S^{q-1})} &=& \frac{2\pi}{\delta^2} + 2 \pi \log \delta + O(1) \ ,   \label{reg_H3} \\
\frac{\partial_I W_{3,q}}{f_I \Vol(S^{q-1})} &=& \frac{2\pi^2}{3 \delta^3} - \frac{5 \pi^2}{3 \delta} + \frac{4\pi^2}{3} + O(\delta) \ .
\ee
The coefficients of the power law divergences are not scheme independent, since they can be modified by counter-terms written on the boundary. 
The possible counter-terms take the schematic form
\begin{align} \label{counterterm1}
\int \d^{p} x \sqrt \gamma \,  \kappa_n(\lambda) R^{n} \Lambda^{p-2n}\, ,
\end{align}
where $R$ is a curvature tensor intrinsic to the defect.  The extrinsic curvature on these equatorial defects vanishes.

For $p$ even, a finite counter-term can alter the constant term in the result, but the coefficient of the log is physical. As we have emphasized above, this is congruous with the fact that the even dimensional boundary can support an ``intrinsic'' $a$-anomaly,
$T^\mu_{\; \mu} = a  E_p \delta^{(q)}(z)+\ldots$. 

We can now show explicitly that the partition function for an equatorial $S^2$ inside $S^d$ matches with our result from the one-point anomaly. Indeed, for a constant coupling $\lambda$ and a constant Weyl transformation with $\sigma = \log \Lambda = - \log \delta$ the anomaly in the partition function \eqref{delta_sigmaW} takes the form
\begin{align} \label{}
\Lambda \partial_\Lambda W_{2,q} = -\frac{a(\lambda)}{4\pi} \int \d^2 x \sqrt \gamma \, R \, .
\end{align}
The intrinsic Ricci scalar for the equatorial defect $S^2$ is $R=2$, which leads to 
\begin{align} \label{}
W_{2,q} = -2 a(\lambda) \log \Lambda \, .
\end{align}
Taking the derivative $\partial_I W_{2,q}$ and using the result from the consistency condition \eqref{farel}
\begin{align} \label{}
\partial_I a =  \pi \Vol(S^{q-1}) \, f_I 
\end{align}
precisely reproduces the logarithmic contribution to \eqref{partialW}.

For $p$ odd, the constant term on the defect is analogous to the constant term $F$ that one gets in odd dimensions without a boundary.  We would like to argue that the real part of this constant term on the defect, let us call it $F_p$, is scheme independent.
Regarding boundary counter terms, there can be no $\Lambda$ independent counter-term on the boundary of the form (\ref{counterterm1}).  However, the 
boundary can support Chern-Simons terms which may alter the imaginary part of $F_p$
 \cite{Closset:2012vg,Closset:2012vp}.
Regarding contributions from the bulk, 
there are bulk analogs of \eqref{counterterm1} which may arbitrarily shift $F_p$ when $d$ is even.
When $d$ is odd, the bulk $F$ will add to $F_p$.
To eliminate these types of ambiguity, we may consider the ratio $|Z_{p,q}|/Z$,
 where $Z_{p,q}$ is the defect partition function and $Z$ is the partition function in the absence of the defect. 
 Here the absolute value removes the imaginary part of $F_p$. 
 Analogously, in the boundary case, ref.\ \cite{Gaiotto:2014gha} proposed to extract the boundary $F_{d-1}$ from the ratio $|Z_{HS^d}|^2/Z_{S^d}$.

$F_p$ defined in this way is expected to have properties similar to $F$ in the bulk as regards to RG flows on the defect. 
However, unlike the $F$ computed in the absence of a defect or boundary, $F_p$ can depend on marginal parameters.
Indeed, it is related to the one-point coefficient by $\partial_I F_p \sim f_I$. 

\vskip 0.1in
\noindent
{\it Discussion and Examples}

There are several explicit examples in the literature where we can observe the marginal coupling dependence
of defect CFT partition functions.  
Perhaps the most well known is the computation of 1/2 BPS Wilson loops in ${\mathcal N}=4$ super Yang-Mills theory
\cite{Erickson:2000af,Drukker:2000rr,Pestun:2007rz}.  The Yang-Mills coupling is exactly marginal.  In our notation, $p=1$ and $q=3$.  The constant term in the partition function, or $F_1$, 
corresponds to the expectation value of these Wilson loops.  
The expectation value can be computed exactly and depends on the coupling in a nontrivial way.  

An example which relates more closely to the $p=2$ case studied in detail here are 1/2 BPS surface operators in
${\mathcal N}=4$ super Yang-Mills. 
A computation in ref.\  \cite{Jensen:2018rxu} shows that the corresponding boundary $a$-anomaly is independent of marginal parameters, at least at leading order in a large 't Hooft coupling expansion.  
However, as shown in \cite{Drukker:2008wr}, the corresponding one-point functions of the marginal operators vanish in this limit as well, consistent with our result.
 
A third related example is a system with a 4d photon coupled to 3d charged matter on a boundary.  In this case, the coupling is exactly marginal and the one-point function for $\langle F_{\mu\nu} F^{\mu\nu} \rangle$ is nonzero (where $F_{\mu\nu}$ is the photon field strength).  As expected, there is a scheme independent constant term on the boundary, $F_3$ in our notation, which depends non-trivially on the coupling \cite{DiPietro:2019hqe}.  
  
A more baroque example involves the position dependent couplings of \cite{Herzog:2019bom}. In that work, partition functions of free fermions and scalars (for $q=1$) were shown to depend explicitly on a ``conformal mass'' parameter.

In the future, it would be interesting to
study in more detail the cases where $p \neq 2$;
explore the constraints that supersymmetry adds to this story;
see if the $b$-type anomalies \cite{Deser:1993yx} can be related in a similar way to the correlation functions of marginal operators.

\noindent

Note added: \cite{Bianchi:2019umv}, which appeared shortly after this letter, comes to similar conclusions about the dependence of the partition function on marginal couplings.

\vskip 0.1in
\noindent
{\bf Acknowledgements:} 
We would like to thank O.~Aharony, D.~Anninos,  F.~Benini, N.~Doroud, N.~Drukker, Z.~Komargodski, E.~Lauria, R.~Rodgers, A.~Schwimmer and M.~Serone for discussion.
C.H.\ was supported in part by the U.K.\ Science \& Technology Facilities Council Grant ST/P000258/1 and by a Wolfson Fellowship from the Royal Society.
I.S.\ is supported in part by the MIUR-SIR grant RBSI1471GJ ``Quantum
Field Theories at Strong Coupling: Exact Computations and
Applications'' and by INFN Iniziativa Specifica ST\&FI.

\end{document}